\documentclass[paper=letter,american,fontsize=11pt,DIV=10]{scrartcl}
\usepackage{times}
\usepackage[T1]{fontenc}
\usepackage[utf8]{inputenc}
\usepackage{hyphenat}
\usepackage{babel}
\usepackage{amsmath}
\usepackage{amssymb}
\usepackage{amsthm}
\usepackage{verbatim}

\usepackage{graphicx}	

\usepackage[authoryear,round]{natbib}
\usepackage{latexsym}
\usepackage{authblk} 
\usepackage[unicode=true,bookmarks=true,bookmarksnumbered=false,bookmarksopen=false, breaklinks=false,pdfborder={0 0 0}, backref=false,colorlinks=false]{hyperref}   

\renewcommand{\vec}{\boldsymbol}

\newtheorem*{theorem*}{Theorem}

\begin{document}
\pagenumbering{roman}

\title{Understanding Physics: \\
`What?', `Why?', and `How?'}
\author{Mario Hubert}
\affil{California Institute of Technology\\
Division of the Humanities and Social Sciences\\
------\\
\emph{Forthcoming in the European Journal for Philosophy of Science}
}

\date{\today}
\maketitle
\begin{abstract}
I want to combine two hitherto largely independent research projects, scientific understanding and mechanistic explanations. Understanding is not only achieved by answering why-questions, that is, by providing scientific explanations, but also by answering what-questions, that is, by providing what I call \emph{scientific descriptions}. Based on this distinction, I develop three forms of understanding: understanding-what, understanding-why, and understanding-how. I argue that understanding-how is a particularly deep form of understanding, because it is based on mechanistic explanations, which answer \emph{why} something happens in virtue of \emph{what} it is made of. I apply the three forms of understanding to two case studies: first, to the historical development of thermodynamics and, second, to the differences between the Clausius and the Boltzmann entropy in explaining thermodynamic processes.
\end{abstract}

\newpage

\tableofcontents

\newpage

\section{Introduction}
\pagenumbering{arabic}
Why do bodies fall to the ground? Why do gases expand in a box? Why do black holes bend light? These are all important scientific questions that demand a \emph{scientific explanation}---indeed, scientific explanations are defined to be answers to why-questions. This definition goes back to the deductive-nomological model \citep{Hempel:1948aa}, in which a scientific explanation is a logical relationship between the explanans, which consists of universal laws and statements concerning particular facts, and the explanandum, which is the event to be explained. 

Similar approaches to the meaning of a scientific explanation can be found in competing theories. When \citet[][Ch. 5.4]{Fraassen:1980aa} describes his pragmatic theory of explanation, he is explicit about the aim of a scientific explanation, ``An explanation is an answer to a why-question. So, a theory of explanation must be a theory of why-questions.'' Here, an adequate answer to a why-question gives context dependent relevant information about the explanandum. A similar restriction of the scope of scientific explanations can be found in Woodward's interventionist account of explanation: ``Instead, I focus on a much narrower range of explanatory activities---roughly, those that consist of providing \emph{causal} explanations ([\dots]) of why some particular outcome or general pattern of outcomes occurs'' \citep[][p.\ 4]{Woodward:2003aa}.

But there is another class of questions that has been mostly overlooked in the literature on scientific explanations. These are what-questions as in the following:  What is gravity? What is a gas? What is a black hole? What is light? I think these kinds of questions have generally been overlooked because the last common ancestor of all modern theories of scientific explanations, the deductive--nomological model, was situated in the logical empiricist tradition, where what-questions were regarded as rather dubious or metaphysical. Although Hempel explicitly distinguished what-questions from why-questions, he only pursued the why in his theory of explanation: ``In times past questions as to the \emph{what} and the \emph{why} of the empirical world were often answered by myths [\dots]'' \citep[][p.\ 9]{Hempel:1962aa}. 

A notable exception to this tradition are mechanistic explanations \citep{Machamer:2000aa,Bechtel:2005aa,Glennan:2017aa, Glennan:2018aa}, where an explanation explains in terms of a mechanism. Before answering \emph{why} a phenomenon happens, a mechanistic explanation presupposes an answer to \emph{what} a system or an object is composed of. I call answers to what-questions, which are the prerequisite for scientific explanations in general, and mechanistic explanations in particular, \emph{scientific descriptions}. Scientific descriptions need to be distinguished from what-questions of a more philosophical kind that are the main topic of metaphysics, epistemology, and ethics  (What is a substance? What is knowledge? What is justice?). I don't aim at a criterion for demarcating these two types of descriptions; instead, I want to argue that scientific descriptions play an important and probably crucial, although underrated, role for \emph{understanding} science, and in particular physics, as why-questions do.\footnote{In \emph{The Open Society and Its Enemies}, \citet[][p.\ 30]{Popper:1945th} criticizes methodological essentialism and defends methodological nominalism, in which what-questions are not important: ``The methodological nominalist will never think that a question like `\emph{What is} energy?' or `\emph{What is} movement?' or `\emph{What is} an atom?' is an important question for physics; but he will attach importance to a question like: `How can the energy of the sun be made useful?' or `How does a planet move?' or 'Under what condition does an atom radiate light?' And to those philosophers who tell him that before having answered the `what is' question he cannot hope to give exact answers to any of the `how' questions, he will reply, if at all, by pointing out that he much prefers that modest degree of exactness which he can achieve by his methods to the pretentious muddle which they have achieved by theirs.'' Obviously, I don't agree with Popper, but he also uses what-questions and how-questions in a more general way than I do.}

There are two main camps in elucidating what the nature of understanding is. For one side, understanding is a mental act of ``grasping'' the object of understanding. In order to know what understanding is, one has to analyze what the act of grasping is by itself and what are the objects one aims at grasping. For \citet{Zagzebski:2001aa}, understanding results from grasping relations among our beliefs in such a way that our mind has conscious transparency about how these beliefs are related. 
The problem with this account is that we may have conscious transparency of why a certain situation occurred and, therefore, we may understand this situation, and still we can be wrong about the states of affairs in the world. In Zagzebski's theory, we can have complete understanding and still fail to match the facts. That is why \citet{Kvanvig:2003aa} adds that understanding must comprise truth \citep[similarly][]{Strevens:2013aa}. 
 
In all these accounts, what is fundamental about understanding is this internal or psychological mental act of grasping. Against this \emph{internalist theory of understanding}, there has arisen an \emph{objectivist theory of understanding}, which has shifted the focus more on analyzing understanding in terms of knowledge and explanation. In this spirit, \citet{Trout:2002aa} attacked the internalist theory: scientists throughout history have been deceived by their internal (subjective) feelings of having understood something, because they can be wrong about the explanation without realizing that they are wrong. So a sense of transparency by itself is neither necessary nor sufficient for understanding. Genuine understanding, on the other hand, hinges on correct (or approximately correct) explanations. 

I will focus on the theory of scientific inquiry and the aims of understanding by \citet{Grimm:2008aa,Grimm:2016aa} and the contextual theory of understanding by \citet{Regt:2017aa} as candidates for objectivist theories, and argue that both approaches still leave out what-questions. Their theories, therefore, not only lack one crucial aspect of scientific understanding, but also allow physical theories that do not have the sufficient resources for scientific descriptions to provide genuine understanding---this impacts other scientific virtues as well. Grimm and de Regt, in general, try to trace how scientists gain understanding in their actual practice, while I aim at discovering norms or criteria for understanding. I am, in particular, interested in exploring why, in certain circumstances, mechanistic explanations provide a deeper sense of understanding. I will be presenting several physical examples  that show how mechanisms provide deeper understanding; in particular, I investigate in detail the influence of mechanisms on the historical development of thermodynamics and on the meaning of entropy. 

Recently, \citet{Illari:2019aa} combined the contextual theory of understanding with mechanistic explanations, hitherto rather independent fields of research, yielding a theory of \emph{mechanistic understanding} in terms of intelligible mechanistic explanations. While she primarily singles out the usability of mechanistic explanations for building models and simulations as the main criteria for mechanistic understanding, I focus on the epistemic merit of mechanistic explanations and argue why understanding in the sense of Grimm and de Regt is, in some cases, not understanding enough.

\section{Scientific Inquiry as Asking `Why?' \emph{and} `What?'}

The first step of my argument is to discuss why we begin with scientific endeavor in the first place and what we aim at when doing science. The epistemic goal we want to reach is scientific \emph{understanding} (for a defense of this view, see also, \citealt[][]{Potochnik:2015aa,Potochnik:2017aa}, whereas \citealt{Elliott:2014aa} show how non-epistemic values influence the practice of science, too), and I argue that one crucial part to reach this goal is to answer what-questions. Whether this is correct depends on what one demands from a theory of understanding. As I show, Grimm's theory of scientific inquiry  leaves out this aspect, while answers to what-questions  played an important role already for Aristotle and also for Wesley Salmon. 

\subsection{The Origin and Epistemic Goals of Scientific Inquiry}
\label{subsec:epist-goal}

Why do certain phenomena  need an explanation? And when is a phenomenon satisfactorily explained? Naively, one would think that a certain phenomenon needs an explanation when we are puzzled or surprised about this phenomenon, and the phenomenon is satisfactorily explained when we cease to be puzzled.
As \citet[][section 2]{Grimm:2008aa} rightly points out, this account of scientific inquiry is insufficient. First, it is circular, because ``a situation would be in need of explanation on this view (hence puzzling) in virtue of its puzzlingness'' (p.\ 484--5). Second, there are many phenomena that require an explanation although these phenomena are not (or no longer) puzzling or surprising. Being puzzled or surprised is too much a subjective feeling as to justify all demands for explanation.  

 \citet[][]{Grimm:2008aa}, therefore, suggests another account, which is inspired by the following quote: 
 
 \begin{quote}
I am tempted to say that explanation locates something in actuality, showing its actual connections with other things, while \emph{understanding} locates it in a network of possibility, showing the connections it would have to other nonactual things or processes. \citep[][p.\ 12]{Nozick:1981aa}
 \end{quote}
The first part describes what an explanation is: a phenomenon is explained when one relates the phenomenon to other things that have really happened.
The second part is crucial for Grimm's theory of explanatory inquiry and understanding because ``\emph{the need for explanation arises for us in the first place only because we view the world in terms of this network of possibility}'' \citep[][p.\ 486]{Grimm:2008aa}. So the reason why we want to explain, for example, the daily rising of the sun is not primarily that we are surprised by this (maybe we are indeed), but because we imagine that it could be different---that it does not rise or that it moves in a different way. 



Two things need to be noted. First, we seek an explanation if we imagine different ways a certain factual situation could have been.  But not all counterfactual situations are interesting \cite[][section 4]{Grimm:2008aa}; we are not curious about why the sun does not turn into a giant chocolate-glazed donut, for instance. The set of possibilities that are relevant for us is constrained by observations and our background knowledge about the phenomenon. Grimm terms these convictions \emph{proto-understanding}. So only the counterfactuals in our proto-understanding of the situation drive us to seek an explanation. Second, it is not enough that an explanation shows \emph{that} something occurred; the explanation has to be rich enough that it also makes clear \emph{why} the factual situation occurred and not the counterfactual. Newtonian mechanics, for instance, can make claims about counterfactuals because it postulates laws of nature.  

There are, however, situations that demand an explanation, although we may not imagine any counterfactuals at all. Furthermore, even if we had the laws of nature and could use them, we may not  completely understand the phenomenon. Grimm is aware of this in a later paper:

\begin{quote}
\label{pg:grimm-visual}
that even though one might possess formulae that describe how the various elements of a system relate to one another, and even though one might be able to manipulate the formulae in order to predict and control how the system will unfold, the system might nonetheless seem like a black box, and thus in an important sense unintelligible. In particular, it is often said that unless one can visualize the system, perhaps by means of a model, the system will remain unintelligible. \citep[][p.\ 219]{Grimm:2016aa}
\end{quote}
It is correct that when one uses laws in an operational way, one does not understand the behavior of the system in an important sense: at least, the system is not visualizable. But the issue runs deeper than just being not visualizable. Those theories that introduce the objects as black boxes do not answer \emph{what a physical  object is made of}. This may be posed independently of any behavior. It is an independent source of scientific inquiry to ask for the inner structure of an object. This type of question is mostly bracketed out in the literature on scientific explanation---probably due to the aftermath of logical positivism. 

\begin{figure}[ht]
\centering
\includegraphics[width=9cm]{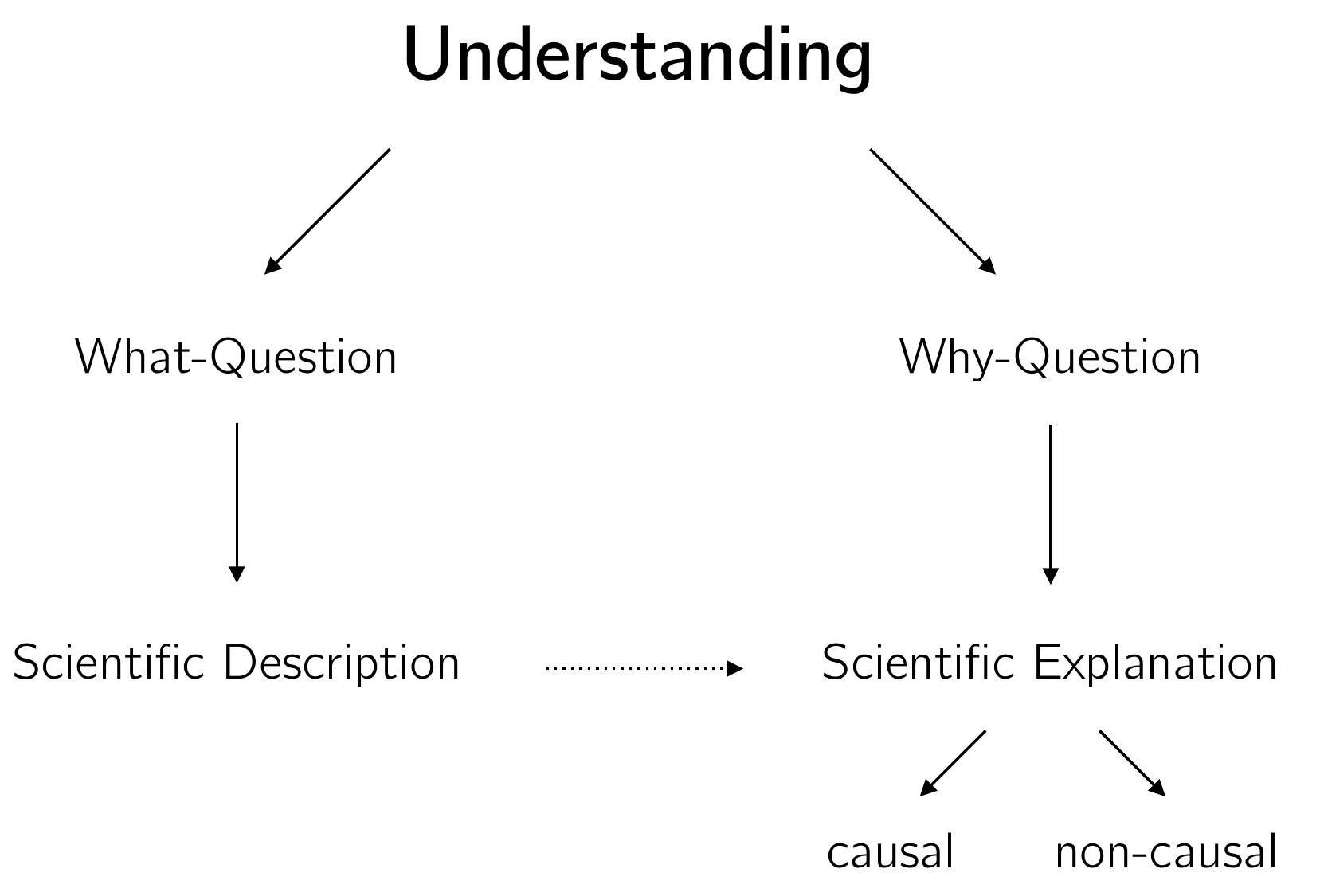}
\caption{Relation between understanding, explanation, and description. Understanding can be gained in two different ways, by asking a what-question or a why-questions. Answers to what-questions are scientific descriptions, and answers to why-questions lead to a scientific explanations. Whereas scientific explanations occur in two different forms, namely, causal or non-causal, there is not yet a further fine-grained taxonomy of scientific descriptions. The dashed arrow indicates that a scientific description may be used (in different degrees) in scientific explanations.}
\label{fig:understanding-diagram}
\end{figure}

An early exception within this tradition, however, is Wesley Salmon, who emphasized both aspects of scientific inquiry:
\begin{quote}
We want to know \emph{how things work} and, it should be added, \emph{what they are made of}. This may be characterized as causal-mechanical understanding [\dots]. It is the kind of understanding we achieve when we take apart an old-fashioned watch, with springs and cogged wheels, and successfully put it back together again, seeing how each part functions in relation to all the others. Before we execute this process, the watch is like a `black box' whose internal workings are mysterious. What we want to do is open up the black box and see how it works. \citep[][p.\ 87]{Salmon:1998ab}
\end{quote}
Here, Salmon explicitly distinguishes why-questions from what-questions, and he says that answers to both kinds of questions give us a particular high form of understanding, in his account a \emph{causal-mechanical understanding}.  It seems that \emph{what something is made of} is for Salmon only an intermediary project in order to be used in a scientific explanation to answer why an event occurred. But what-question can be independently posed from a why-question and has a value to be answered by itself, although I agree with Salmon that the insights from such an answer can be used in a scientific explanation. In order to distinguish scientific explanations from answers to what-questions, I will call the latter \emph{scientific descriptions} (see Fig.\ \ref{fig:understanding-diagram}).\footnote{I can envision a rough taxonomy for scientific descriptions into material, structural, and functional descriptions, but such a project is still undone and would be the subject of another paper.}

A certain class of causal explanations along the lines envisioned by Salmon are \emph{mechanistic explanations}, which have been thoroughly investigated by philosophers in the last two decades initiated by \citet{Glennan:1996aa} and \cite{Machamer:2000aa}. I give a more thorough overview and detailed discussion of mechanistic explanation in the next section. For now it suffices to know that mechanistic explanations start with answering what a physical system (or, in general, a mechanism) is composed of. The parts of the physical system that are identified are used to explain why a certain phenomenon exemplified by the physical system occurs. Mechanistic explanations answer two questions about a phenomenon instead of one and explain \emph{how} a phenomenon occurs in virtue of the behavior of its constituent parts. Therefore, mechanistic explanations are particularly epistemically precious. They provide a deeper understanding than an explanation that merely focuses on the `Why?'.  I, therefore, distinguish three kinds of understanding:
\begin{enumerate}
\label{pg:understanding-kinds}
\item
\emph{Understanding-what}: The form of understanding of a thing or an object provided by a scientific description.
\item
\emph{Understanding-why}: The form of understanding of a phenomenon provided by a scientific explanation. 
\item
\emph{Understanding-how}: The form of understanding of a phenomenon provided by a scientific explanation that explains in virtue of the behavior of the parts identified by a scientific description. 
\end{enumerate}
It is possible to gain a certain form of understanding, namely, understanding-what, even without a scientific explanation. A scientific description gives information about the constitution of a physical object or a physical system, and, therefore, provides some form of understanding of this object or system. Similarly, one can have a certain form of understanding, namely, understanding-why, without having a scientific description or with an incomplete or partial scientific description. Understanding-how, on the other hand, combines (aspects of) understanding-what and understanding\hyp{}why by explaining why something comes about in terms of its constituents. Since more questions about a phenomenon are answered, understanding-how is a deeper form of understanding than the other two. I don't think it's reasonable to make a hierarchy between understanding-what and understanding-why, because these are different kinds of questions aiming at different aspects of a phenomenon. 

Recently, \citet{Illari:2019aa} analyzed how mechanistic explanations provide understanding, too, but her approach is more pragmatic:
\begin{quote}
A phenomenon is mechanistically understood when scientists have an intelligible
me\-chanistic explanation for the phenomenon; i.e. a mechanistic explanation that they
can use. \citep[][p.\ 69]{Illari:2019aa}
\end{quote}
 In contrast to my account, which focuses on how much we can learn or answer about the world, Illari identifies mechanical explanations as \emph{useful} for building models and simulations without evaluating whether mechanical explanations yield a better or deeper kind of understanding. Nevertheless, she mentions that ``[u]nderstanding comes in greater and lesser degrees'' (p. 82), but a hierarchy of understanding still needs to be worked out \citep[a promising path is shown in][]{Kelp:2015aa}. I agree with Illari that mechanistic explanations are useful in practice, but that is not the reason, for me, that they yield understanding---it is rather a pragmatic fallout. In this sense, I side more with Salmon's approach: there is something intrinsic about mechanistic explanations that they provide understanding, namely, by means of a scientific description.

\subsection{Scientific Descriptions and Aristotle's Four Causes}

I now want to compare this scheme of understanding with Aristotle's famous theory of the four ``causes'',\footnote{The Greek word that Aristotle uses is $\alpha \iota \tau \iota \alpha$ (\emph{aitia}), which is standardly translated as \emph{cause}. Since \citet{Vlastos:1969aa} distinguishes between Aristotle's \emph{causes} and \emph{becauses}, there is a debate whether to translate $\alpha \iota \tau \iota \alpha$ rather as \emph{explanation} \citep[see also][section III for a defense of \emph{explanation}]{Annas:1982aa}. We will see in the following that \emph{explanation} would be more suitable in my scheme.} one of the first systematic treatments of this topic,\footnote{Probably the first extant discussion of understanding and explanation is Plato's \emph{Phaedo}, which Aristotle explicitly criticizes \citep{Annas:1982aa}.} because \citet{Grimm:2008aa} begins his paper with Aristotle's idea for the origin of scientific inquiry and because, I think, Aristotle's theory of the four causes is still insightful for current debates, especially for presenting different kinds of scientific descriptions.
 
Aristotle's primary question about scientific inquiry is similar to Grimm's: ``For the point of our investigation is to acquire knowledge, and a prerequisite for knowing anything is understanding \emph{why} it is as it is---in other words, grasping its primary cause'' \citep[Physics, II.3, 194b16--b22, translated in][p.\ 38--9]{Aristotle:2008aa}. Right after this sentence, Aristotle lists the four causes, material, formal, efficient, and final, without justifying where they come from or why they are complete \citep[][p.\ 701]{Stein:2011aa}. This lacuna by Aristotle has been filled by \citet{Hennig:2009aa}, whose account is relevant for my interpretation:\footnote{As Hennig mentions in section VI by citing Avicenna and Aquinas, there have been many attempts to unify the four causes in the history of philosophy.} 

\begin{figure}[ht]
\centering
\includegraphics[width=15cm]{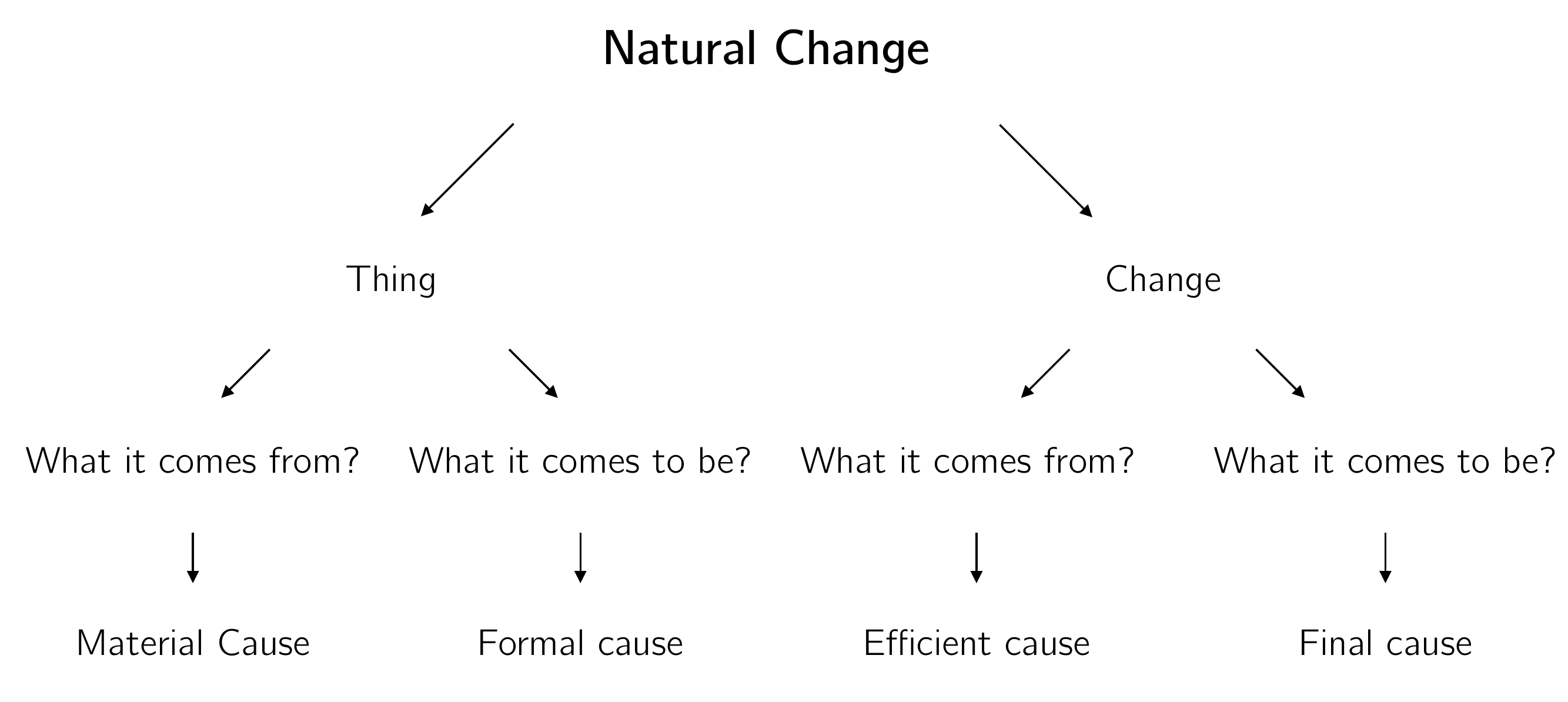}
\caption{Hennig's scheme of the four causes of Aristotle. Natural change is decomposed into the thing undergoing the change and change itself. Of the thing and change, two identical pairs of questions can be asked: `What it comes from?' and `What it comes to be?'. By answering these questions for the thing and for change, it turns out that what the material cause is for the thing is the efficient cause for change, and what the formal cause is for the thing is the final cause for change.}
\label{fig:aristotle-hennig-diagram}
\end{figure}

\begin{quote}
My aim is to say what Aristotle said, not in the sense of reporting what he said, but rather in the sense of repeating and elaborating what he said. I will show that his fourfold distinction naturally arises from the combination of two distinctions that apply to all natural beings. First, concerning any natural change, we may distinguish between the thing that changes and the change that it undergoes. Neither of these could be studied without in any sense referring to the other. Second, we may ask out of what a natural thing comes to be what it is, and we may ask an analogous question about a natural change. Conversely, we may ask what a thing or a change naturally comes to be. Again, the two questions, out of what something comes to be what it is, and what it comes to be, cannot be separated. \citep[][pp.\ 137--8]{Hennig:2009aa}
\end{quote}
In Hennig's interpretation, Aristotle focuses his account on natural change, which is defined to be a typical or regular change or pattern in nature \citep[][section I]{Hennig:2009aa}. And this natural change has two components: first, a thing that changes, and second change itself (see Fig.\ \ref{fig:aristotle-hennig-diagram}). For example, the rolling of a billiard ball on a slope consists of the billiard ball, the thing undergoing the change, and the motion of the billiard ball. Next, we can ask two questions of the thing and change: (i) `Out of what does a thing or change come to be what it is?' and (ii) `What does a thing or change come to be?' 

Hennig argues that (i) is answered by the material cause for a thing and the efficient cause for change, whereas (ii) is answered by the formal cause for a thing and the final cause for change (Fig.\ \ref{fig:aristotle-hennig-diagram}). A comprehensive justification for these answers demands some familiarity with Aristotle's metaphysics, especially the role of actuality and potentiality, whose thorough examination would go beyond the scope of this paper, so I briefly illustrate Hennig's scheme by examples. The material cause of a billiard ball is what it consists of, that is, its matter.\footnote{\citet[][section II]{Hennig:2009aa} emphasizes that Aristotle has something more general in mind than what something consists of, namely, \emph{that which potentially is the result}; therefore, there are other examples where the material cause is not only matter or not matter at all.} The formal cause of a billiard ball, that what comes to be, is the billiard ball---it remains to be a billiard ball. But, if the material cause of the billiard ball were to be radioactive matter, the formal cause, that into which the billiard ball typically changes after some time, would be a new atomic composition plus radiation. The efficient cause and the final cause do not, according to Hennig, characterize a thing but change itself. The efficient cause of change is not only the cause that initiates the change but that keeps the change to continue \citep[][section IV]{Hennig:2009aa}. The efficient cause of the motion of a billiard ball on the table, hence, is not only the strike of a cue, but also the inertia of the ball (if the ball keeps moving). The final cause of change is in Hennig's interpretation the end or the direction of change. As in the case of the efficient cause, the final cause is not only to be construed as a final stage or result, but it can also come or come into being in a certain chain of steps; it is ``primarily the `limit' or standard pattern to which [change] proceeds'' \citep[][pp.\ 152--3]{Hennig:2009aa}. The final cause of a falling billiard ball, accordingly, would be to hit the ground, or the final cause of a radioactive decay would be to result in radiation. 

I don't intend to criticize Henning's account of Aristotle's four causes, although it seems a bit unsatisfactory or confusing that the material and formal cause of a thing also refer to change, although change is separated from the thing, but this may be rather due to Aristotle's idiosyncratic view of potentiality; instead, I want to give a complementary, epistemic interpretation of the four causes in terms of explanations and descriptions---in doing so I depart from the historical Aristotle and focus only on physics, unlike Aristotle, who aimed at a scheme for change in general, including, biology and human action. Hennig presents an ontological view of the four causes by referring to the things and the changes and processes in the world by starting to dissect what natural change is, and that may be the reason why he settles for \emph{cause} as the translation of \emph{aitia} instead of \emph{explanation}.\footnote{\citet{Cartwright:2020aa} distinguish two questions, which show the difference between my and Hennig's approach: 

\begin{enumerate}
\item
An epistemological question: “What kind of explanation is involved?”
\item
An ontological question: “What is going on \emph{in the world}?”
\end{enumerate}

} 

\begin{figure}[ht]
\centering
\includegraphics[width=9cm]{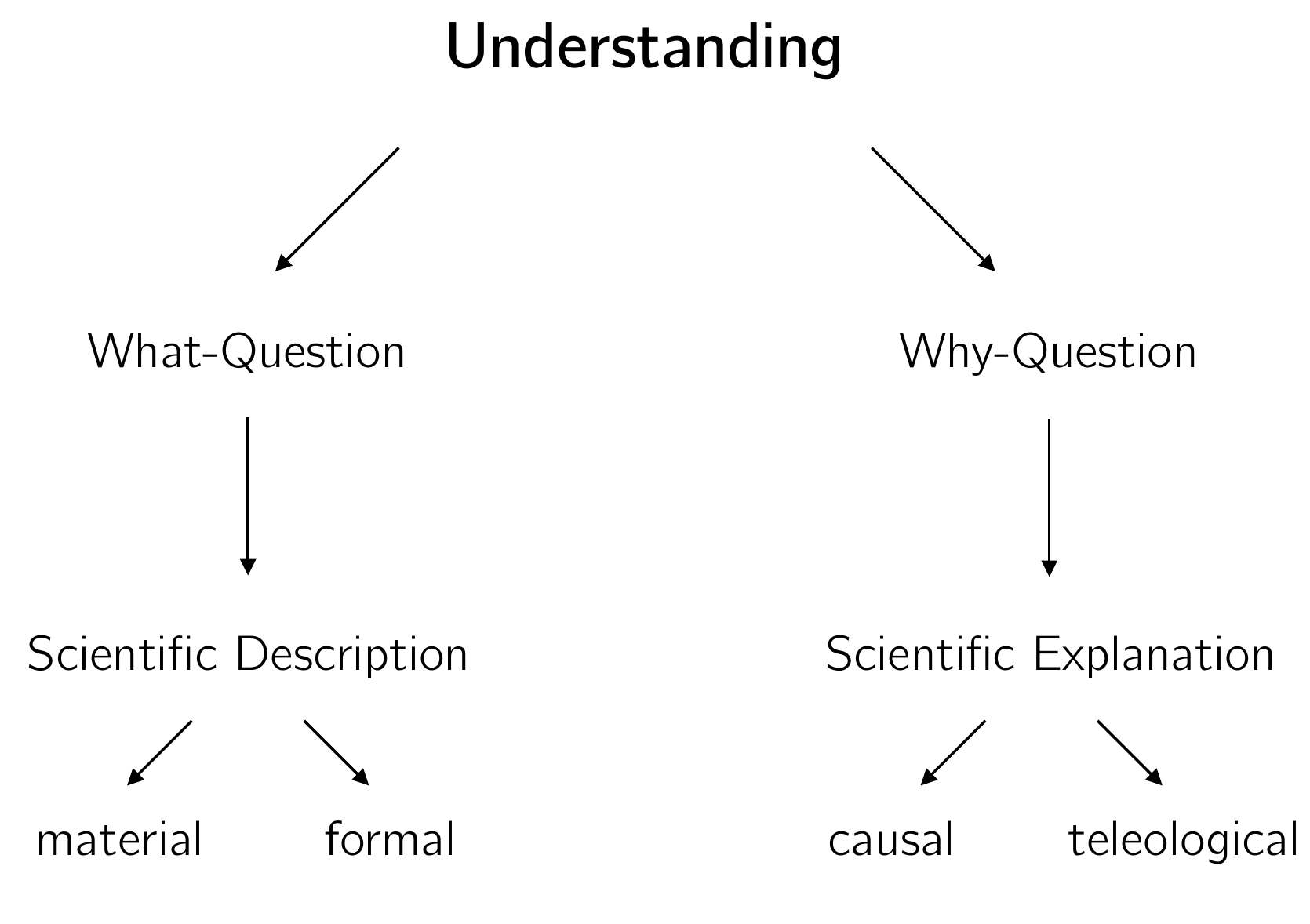}
\caption{An alternative interpretation of Aristotle's four causes.}
\label{fig:Aristotle-understanding}
\end{figure}

If we take, however, the view of, for example, \citet{Vlastos:1969aa}, \citet{Annas:1982aa}, and \citet{Stein:2011aa} that \emph{aitia} means explanation (or something similar to explanation), we can start our investigation with our requirements for understanding. 
Let us look again at Aristotle's quote on how to gain knowledge or understanding:
``For the point of our investigation is to acquire knowledge, and a prerequisite for knowing anything is understanding \emph{why} it is as it is---in other words, grasping its primary cause'' \citep[Physics, II.3, 194b16--b22, translated in][p.\ 38--9]{Aristotle:2008aa}. The basic question (and it seems the only basic question to acquire knowledge) is to ask `Why?' which is answered by an explanation. I think, however, that Aristotle's formulation of ``\emph{why} it is as it is'' incorporates two questions: \footnote{I don't claim that this is what Aristotle indeed had in mind, but rather that this is a plausible way to interpret the passage.} there is a why-question about change or phenomena, but there is also a what-question about things or physical objects. And in contrast to Hennig, these questions arise from epistemic requirements about the nature of understanding, which, in my view, incorporates these two aspects. There are two ways to answer the what-question in Aristotle's scheme, and two ways to answer the why-question (see Fig.\ \ref{fig:Aristotle-understanding}). Therefore, the question `What is the billiard ball made of?'' can be answered by two kinds of descriptions: the actual matter of the billiard ball (material cause) or the structure of the matter parts (formal cause) (to be close to Hennig's scheme, the formal description of the billiard ball can also be a description of the potential form of the billiard ball). `Why does the billiard ball move the way it does?' is answered by a scientific explanation, and for Aristotle this would be a causal or a teleological explanation. 

We, thus, see that my approach from understanding leads to the same pairing of causes, that is, material/formal and efficient/final, as Hennig's scheme. Material and formal causes or descriptions are answers to what-questions about things or physical objects, while efficient and final causes or explanations are answers to why-questions about change or physical phenomena. 

\citet{Greco:2014aa} also proposes an update to Aristotle's four causes. He proposes, which I agree on, that it is more suitable to translate the Greek \emph{episteme} as \emph{understanding}, instead of the traditional \emph{knowledge}. Aristotle, according to Greco, proposed a scheme for how to gain understanding, namely, by giving an explanation or an account. Citing one of the four causes counts as such an account. In Greco's neo-Aristotelian update, the four causes exemplify four different kinds of dependence relations:
\begin{quote}
[W]hat do Aristotle's four causes have in common? One way to think of it is that they each cite some kind of `dependence relation.' \citep[][p.\ 290]{Greco:2014aa}
\end{quote}
Introducing dependence relations make it particularly straightforward to generalize Aristotle's scheme to modern dependence relations: (i) efficient causal relations, (ii) constitutive (`material') relations, (iii) essential (`formal') relations, and (iv) teleological relations. Greco adds further relations, like (v) mereological relations, (vi) logical or mathematical relations, (vii) conceptual relations, and (viii) supervenience relations. For Greco, however, all these dependence relations are on a par, because they are used to answer why-questions. But causal relations are of a different kind than mereological relations relations, for example. Causal relations explain \emph{why} some event comes about due to its relation in a causal network, whereas mereological relations describe \emph{what} something is made of by citing how parts are related to wholes.

\section{Scientific Descriptions and Mechanisms}

I have argued that we not only gain understanding by explaining why something happened as it happened and not otherwise, we may gain understanding from the description of the inner structure of an object just for the sake of understanding what it is. Often the inner structure helps us to understand the behavior of the entire object in a mechanistic way. And there are many what-questions in physics that are important by themselves: What is space? What is time? What is motion? What is a black hole? Different what-questions lead to different scientific descriptions. It would be an important project to find a simple modern taxonomy of scientific descriptions as has been found for scientific explanations. An overarching theory has been developed by philosophers to answer ``Why?'' in terms of ``What?'': mechanistic explanations.

This combination of what and why is often subsumed in asking `How?', as a mechanistic explanation explains by means of \emph{how} the constituents of a system generate the phenomenon \citep{Glennan:2017aa}. More precisely, a mechanism is minimally defined in the following way: 
\begin{quote}
A mechanism for a phenomenon consists of entities (or parts) whose activities and
interactions are organized so as to be responsible for the phenomenon. \citep[][p.\ 17]{Glennan:2017aa}\footnote{There are two other definitions of mechanisms often discussed, one presented by \citet{Machamer:2000aa} and the other by \citet{Bechtel:2005aa}, which are rather tailored to the biological sciences than to physics.}
\end{quote}
This definition illustrates in which way an answer to `How?' is decomposed into an answer to `What?' and `Why?'. 

First, it is said that one needs to give a description of the parts of a mechanism, that is, the constituents that give rise to a phenomenon. It is assumed that these parts are discrete \citep[][p.\ 339]{Kuhlmann:2014aa} and that they have a location in space \citep[][p.\ 34]{Glennan:2017aa}. From the examples of mechanism, I take it, the space is the three-dimensional space that we perceive, although I can envision to generalize to more complicated spaces in order to have mechanisms for modern physics, like string theory or quantum gravity. There are two ways to understand what it means for the parts of a mechanism to be discrete. Either one adheres to a form of atomism, where the parts behave like discrete atoms, or, more generally, the parts can be identified in bounded regions of space, like the parts of the electromagnetic field. \citet[][p.\ 34]{Glennan:2017aa} seems to imply the latter more general reading, ``Entities can have
locations [\dots] even if they are widely spread out and overlap other
entities. There is no particular limit on how far an entity can be sprawled across spacetime.''

Given a mechanism, a further question is to find out whether it is fundamental or compound, that is, made out of further mechanisms  \citep[][Section 7.5]{Glennan:2017aa}. 
\citet[][Ch.\ 5.1]{Machamer:2000aa}  say that when using a theory one postulates certain fundamental building blocks in the \emph{domain} of the theory, whose existence and properties are taken as given. These are the fundamental parts of a mechanism. 

Second, the why-part of mechanisms, the way it explains a phenomenon, says that the parts of the mechanism interact in an organized way. In other words, the parts of a mechanism stand in causal relations to one another, which are themselves mediated by mechanisms. \citet[][p.\ 145]{Glennan:2017aa} describes the general idea in this way, ``If one event causes a second,
there exists a mechanism by which the first event contributes to the production of the
second. If some kind of event causes another kind of event, it is because there are one
or more kinds of mechanisms by which events of one kind cause events of the second.'' The details of this rough descriptions need to be spelled out. Even if a particular theory of causation is not advocated, it is assumed that the causal relations in a mechanistic explanation are to be \emph{local} \citep{Illari:2011ab}. Locality, in this case, means that the mechanism is closely located to the phenomenon, or, in other words, that the mechanism bringing about the phenomenon is in the approximate spatiotemporal region as the phenomenon itself. This locality requirement, as well as the former requirement of spatiotemporal localization, seems to be challenged by quantum mechanics.

\section{A Critique of Current Theories of Understanding}
\label{sec:understanding-theories}

We will see in this section that current theories of understanding are not so much concerned about  what-questions and ontology, either, but rather trace how physicists, and scientists in general, gain understanding without making normative claims about how physics can be and maybe ought to be done. This does not only miss a crucial aspect of scientific inquiry but also sends a message to physicists that their current practice of only saving the phenomena is good enough, which I oppose. I will reject Grimm's theory that understanding physics is understanding-by-grasping-structures, as well as the de Regt--Dieks theory. Both these theories cannot account for the deeper level of understanding that a theory with a mechanism offers. 

\subsection{Grimm's Account}


\citet{Grimm:2016aa} distinguishes different kinds of understanding in the natural sciences. There is \emph{understanding-as-grasping-of-structure}, which is a pragmatic way to understand nature. It suffices that the theory offers a recipe or algorithm that specifies how manipulating one part changes the behavior of another. Grimm grants, however, that there is a deeper way of understanding, namely, \emph{understanding-as-picturability}. This kind of understanding supplements  \emph{understanding-as-grasping-structure} by making the physical processes in some (unspecified) way visualizable. But as I said in section \ref{subsec:epist-goal}, visualizability is not the real issue a scientific explanation aims at. If visualizable theories want to tell us something profound about nature beyond a heuristic or mnemonic story (and beyond empirical behavior), they need to introduce a mechanism. Otherwise, one may run into a paradoxical situation as in Feynman diagrams. These diagrams make certain quantum processes visualizable, but they are not intended to represent actual microscopic processes in nature  \citep[at least not in the standard interpretation of these diagrams, see][]{Kaiser:2005aa}. If understanding were only to be understanding-as-picturability, Feynman diagrams would give us sufficient understanding of phenomena involving particle creation and annihilation. Therefore, understanding is not only about grasping the structure of the world in an operational way or about making the theory visualizable. 


\citet[][p.\ 220]{Grimm:2016aa} says that \emph{understanding-as-grasping-structure} is ``understanding enough'' in physics.\footnote{Grimm does so in order to  argue that, like in physics, \emph{understanding-as-grasping-structure} is understanding enough for understanding people.} His argument is that over time, physicists got more familiar and more comfortable with just grasping structure, that is, with using theories for mere manipulations and predictions, that this kind of understanding turned out to be the kind of understanding science really needs to aim at. For instance, although Newton and his peers felt uneasy with action-at-a-distance, future generations got used to Newton's theory of gravitation and accepted it as a full-fledged physical theory. Newton did not feign any hypothesis about the underlying mechanism of gravitation, because he was not able to find one, \emph{although he kept trying}. His successors, on the other hand, did not think the quest for a mechanism a task worthwhile for physics at all. That is historically correct (at least for gravity\footnote{The situation for electromagnetism looked a bit different: especially in Britain, physicists early on presupposed a mechanism for the electromagentic field, of which Maxwell's vortex model was a prominent example \citep{Siegel:1991aa,Siegel:2014aa}, while in Germany Wilhelm Weber's action-at-a-distance theory was popular until it was challenged by Helmholtz and Hertz at the end of the 19\textsuperscript{th} century \citep{Steinle:2013aa,Buchwald:2013ad}.}), because physicists were more convinced by positivism after Newton and, of course, by the huge experimental success of Newtonian mechanics in accounting for novel phenomena, like the motion of comets, such that a search for an ontological underpinning of gravity seemed to be unnecessary.

Ironically, progress in gravity was finally made by Einstein, for whom grasping structure was not understanding enough \citep{Howard:2005aa}. It is often said that in the early development of special relativity, grasping structure was enough for Einstein, because he seemed to be at this time more sympathetic to positivism. But upon further scrutiny this is incorrect \citep{Norton:2016aa}. Even in his 1905 paper \emph{On the Electrodynamics of Moving Bodies},  Einstein presents a thought experiment, which illustrates that from different reference systems one gets different electromagnetic explanations for the same physical effects \citep{Einstein:1905kx}. Although the logical positivists praised Einstein's axiomatic approach to special relativity, the 1905 paper already shows that Einstein was not only concerned about saving the phenomena but also about certain theoretical virtues of a physical theory, something that became more and more important to him \citep{Giovanelli:2013aa}. Einstein starts the paper by pointing to a unsatisfactory asymmetry of Maxwell's theory of electrodynamics in explaining the current in a piece of metal---it seems that no one before Einstein was deeply concerned about this asymmetry. If a piece of metal is moved through a magnetic field of a resting magnet, a current is induced in the metal by the magnetic field, and if the metal is at rest and the magnet is moving the same current is induced this time, however, by an \emph{electric} field, although the metal and the magnet have the same relative velocity. 

One may think that Einstein's attitude was positivistic and that understanding-as\hyp{}grasping\hyp{}structure was enough for him, because he seemed to argue that the same empirical behavior (that is, the same relative velocity) shall be explained in the same way; in particular, the different explanations ought not to differ in the  behavior of unobservable objects (here, the electric or magnetic field). This example, rather, shows that Einstein was driven, even if he may not be aware of, by trying to \emph{understand} why nature would have this kind of asymmetry and whether it is possible to \emph{explain} the same empirical observation \emph{without} such asymmetry between the electric and magnetic field. A positivist would rather shrug her shoulders and re-interpret the asymmetry and the electric and magnetic field as mere theoretical artifacts of  electromagnetism---only the empirical predictions would matter. Einstein, on the other hand, was bothered by this asymmetry because he was not content with the ways relative velocity is explained. He was bothered by the asymmetry and tried to overcome this problem because he was \emph{not} a positivist and because understanding-as\hyp{}grasping\hyp{}structure was not enough for him.

Later, Einstein took up the same basic idea from his 1905 paper, which lead to his general theory of relativity: although there seems to be a theoretical difference between inertial and gravitational mass, it is impossible to ''observationally distinguish between a state of uniform acceleration and the presence of a gravitational field." \citep[][p.\ 624]{Smeenk:2007aa}. Einstein used this equivalence principle to build up a more unified theory of gravity in which the \emph{entire} physics in free falling systems is equivalent to the physics in unaccelerated systems \citep{Norton:1993aa,Hoefer:1994ud}.\footnote{Max Abraham and Gustav Mie pursued a different strategy to build a unified theory of gravity by implementing gravity directly into special relativity \citep{Norton:2007aa,Smeenk:2007aa}. This project was finally abandoned for theoretical and empirical reasons. I thank Dennis Lehmkuhl for this example.}  


Another example Grimm puts forward to support his claim about understanding enough is  quantum mechanics, when Wolfgang Pauli said in notoriously demeaning manner:

\begin{quote}
Even though the demand of these children for Anschaulichkeit [visualizability] is partly a legitimate and a healthy one, still this demand should never count in physics as an argument for the retention of fixed conceptual systems. Once the new conceptual systems are settled, then also these will be anschaulich [visualizable]. \citep[from][p.\ 385]{Regt:2014aa}
\end{quote}
Pauli makes it easy and just redefines what ``anschaulich'' can mean over time. That physicists get used to certain practices only tells us something about the psychology of physicists and their appreciation for the history of their discipline. It is a sociological fact that physicists have gotten more familiar and content with \emph{understanding-as-grasping-structure}; it does not mean that physics has to be so or that it was always so. It is a historically interesting observation that for the development of quantum mechanics grasping structure was regarded as enough, while for relativity theory, which was developed around the same time, grasping structure was \emph{not} enough. This was one of the main points (apart from the issue of non-locality), where Einstein disagreed with Bohr and especially with Heisenberg \citep{Fine:1993um, Beller:1999aa,Becker:2018aa,Strien:2020aa}:

\begin{quote}
Naturally one cannot do justice to [the argument] by means of a wave function. Thus I incline to the opinion that the wave function does not (completely) describe what is real, but only a to us empirically accessible maximal knowledge regarding that which really exists [. . .] This is what I mean when I advance the view that quantum mechanics gives an incomplete description of the real state of affairs. \citep[Einstein in a letter to Paul Epstein on November 10, 1945, quoted in][p,\ 151]{Harrigan:2010aa}
\end{quote}
Einstein is famous for interpreting the wave-function as representing the statistical patterns of an ensemble of identically prepared systems. He granted that for practical purposes the wave-function gives in this way the correct empirical predictions, but quantum mechanics would still lack to explain certain facts about the world. Therefore, quantum mechanics would not give us enough understanding about the goings-on in the world according to Einstein. So, for certain practical purposes understanding structure is understanding enough, but physics in general ought not to end there. 

\subsection{De Regt's Contextual Theory of Understanding}

This pragmatic attitude within philosophy of science can also be observed in the de-Regt--Dieks contextual account of understanding \citep{Regt:2005aa,Regt:2017aa}. This theory is based on two doctrines, the \emph{Criterion for Understanding Phenomena} (CUP) and the Criterion for the Intelligibility of Theories (CIT\textsubscript{1}):

\begin{itemize}
\item[-]
\textbf{CUP}: A phenomenon P is understood scientifically if and only if there is an explanation of P that is based on an intelligible theory T and conforms to the basic epistemic values of empirical adequacy and internal consistency. \citep[][p.\ 92]{Regt:2017aa}
\item[-]
\textbf{CIT\textsubscript{1}}: A scientific theory $T$ (in one or more of its representations) is intelligible for scientists (in context $C$) if they can recognize qualitatively characteristic consequences of $T$ without performing exact calculations. \citep[][p.\ 102]{Regt:2017aa}
\end{itemize}
Understanding is here defined with respect to a scientific \emph{theory}, which itself has to fulfill certain standards: the theory has to be empirically adequate, internally consistent, and intelligible. Intelligibility is here not defined in terms of visualizability but in a pragmatic way: a theory is said to be intelligible if a scientist can easily apply the theory for practical matters \citep[][p.\ 40 and p.\ 101]{Regt:2017aa}, of which CIT\textsubscript{1} is an instantiation particularly suited for physics. Because this kind of intelligibility of a theory does not only depend on the theory itself but also on the abilities and knowledge of the scientist, and maybe on the current state of technology for applying the theory, intelligibility is a non-intrinsic, contextual property of a theory---hence the name of this theory of understanding. Intelligibility is also relative to a particular scientist; so a theory that is intelligible to one scientist may be unintelligible to another. 

The main problem I see with de Regt's account, as with Grimm's, is that it grants too many theories to provide for genuine understanding. In particular, textbook quantum mechanics would give a physicist understanding of the corresponding phenomena. In order to justify the intelligibility criterion, \citet[][pp.\ 101--2]{Regt:2017aa} compares a physical theory with an oracle that happen to give the same correct predictions as the physical theory. The oracle is inferior to the physical theory because it does not give us understanding of how the phenomena are generated; it is rather like a black box that spits out the correct answers. It is obvious that an oracle would not be a satisfying device for scientists: 

\begin{quote}
Scientists want more than this: in addition they want insight, and therefore they need to open the black box and consider the workings of the theory that generates the predictions. Whatever this theory looks like, it should not be merely another black box producing the empirically adequate descriptions and predictions (on pain of infinite regress). In contrast to an oracle, a scientific theory should be intelligible: the theory itself should be understood. The opacity of an oracle entails that the way in which its predictions are generated is unclear. The intelligibility of scientific theory, by contrast, implies that it should be possible to grasp how its predictions are generated. A sign that such a grasp has been achieved is the fact that one has developed a ``feeling'' for the consequences of the theory in concrete situations [\dots]. 
\citep[][pp.\ 101--2]{Regt:2017aa}
\end{quote}
But textbook quantum mechanics  treats the physical systems under considerations as black boxes, although they are intelligible according to de Regt's criterion!  Quantum mechanics is a candidate for a fundamental theory and ought not to leave these issues open. The recipe of quantum mechanics is therefore similar to an oracle, an oracle that gives us a glimpse, though, of the inner goings-on but is secretive about the deeper detailed story \citep[see, for instance,][for this kind of critique of quantum mechanics]{Maudlin:2019aa}. Really opening the black box of quantum phenomena would require a scientific description of the constituents of these systems and how the behavior of these constituents generates the observable phenomena. A highly-skilled physicists having perfect ``feeling'' for the consequences of quantum mechanics would still fail to understand quantum phenomena in the deeper sense.

Yet, one may argue that textbook quantum mechanics in fact differs from an oracle, because quantum mechanics makes predictions by applying certain steps and conceptual tools.\footnote{Thanks to an anonymous reviewer for raising this concern.} In contrast, an oracle just delivers an answer without showing how this answer has been generated. I agree that quantum mechanics does better than an oracle in justifying predictions, but it is still mysterious and unexplained what quantum systems are made of, what the wave-function is, what non-locality is, and what measurement devices are \citep[see also][who identifies three problems of textbook quantum mechanics: the ontology problem, the locality problem, and the measurement problem]{Norsen:2017aa}. All these problems ultimately arise because quantum mechanics does not sufficiently answer \emph{what} a quantum system is made of. This is one major motivation for alternative theories, like the de Broglie--Bohm theory, the GRW collapse theory (which has been even developed with two kinds of scientific descriptions of quantum objects), and Everett's many-worlds theory.

\citet{Regt:2016aa} have made the contextual theory even more pragmatic and turned it into a theory of \emph{effectiveness}. They no longer define understanding relative to a theory but relative to a \emph{representational device},   which can be ``theories, models, and diagrams'' (p.\ 50). The representational device need \emph{not} be internally consistent, but only intelligible (in the above sense of easily usable) and reliably successful. A theory can be reliably successful in three different ways: by making correct predictions (formerly dubbed empirical adequacy), by guiding practical applications, and by developing better science (p.\ 56). Finally, a representational device is said to be \emph{effective} if it is intelligible and reliably successful. Hence, the effectiveness account of understanding goes like this:
\begin{quotation}
According to our \emph{effectiveness condition on understanding}, understanding can only be gained from representational devices that are, for a subject S in a context C, effective in the sense defined above. \citep[][p.\ 55]{Regt:2016aa}
\end{quotation}



As in the contextual theory, the effectiveness theory approves too many theories to provide (genuine) understanding---it's not my concern in this paper to evaluate representational devices other than theories. Are there other options for philosophers to analyze understanding in physics? In particular, are philosophers in a position to criticize the practice of physicists (and scientists in general) and to set norms for what counts as genuine understanding? Although I assume that many would see that as a very unpalatable claim for philosophers to make, philosophers have, to my mind, the proper training and the appropriate tools to do so. First, philosophers have a certain distance and impartiality to physical theories because they are normally not invested in a particular physical theory or a particular physical research project. Second, philosophers of science are trained in evaluating the methodology of science (and physics). Third, philosophers of science are trained in comparing historical case studies, from which certain lessons about the practice of science can be drawn. This, in conjunction with expertise in ontology, epistemology, and often even training in a special science, makes a philosopher qualified to contribute to the debate about scientific norms.

\citet[][p.\ 333]{Regt:2019aa} distinguishes two kinds of norms in a theory of understanding:
\begin{itemize}
\item[-]
\emph{Prescriptive norms,} which provide rules that scientists should follow if they want to achieve scientific understanding of phenomena.
\item[-]
\emph{Evaluative norms,} which can be used to assess whether or not a particular scientific result does in fact provide genuine understanding.
\end{itemize}
As de Regt discusses in the paper, his theory of understanding provides evaluative norms without posing ``\emph{universal} restrictions on theories or theory choice, but it does entail that certain theories are preferable over others in certain \emph{local} context." \citep[][pp.\ 341--2]{Regt:2019aa}. The contextual theory of understanding can, therefore, be used to analyze in which way a certain physical theory yields understanding, but it doesn't say whether this kind of understanding can or ought to be improved on. Basically, it suffices, especially in the effectiveness theory, when physicists successfully apply their theory, even only to build computers and smartphones, that they have (genuine) understanding. But being able to build complex technology or making successful empirical predictions or even to build new theories (the three ways a theory is said to be reliably successful, according to de Regt and Gijsbers) can be possible without knowing the scientific descriptions of the relevant objects. Therefore, one needs prescriptive norms, like the demand for these descriptions, to do physics, otherwise understanding is rendered by whatever scientists do successfully. And how many scientists are needed to qualify a scientific theory to be intelligible or effective? If there are two competitive theories in the same domain, it seems that de Regt's theory cannot distinguish either theory to provide more or less understanding than the other, as long as there is a group (?) of scientists successfully using the theory. The prescriptive norms that I intend to establish for physical theories are not arbitrary. They are, first, motivated by the principal epistemic requirements of understanding and the the basic questions we can pose in our scientific inquiry, and secondly by their historical success and usefulness in different physical theories.

\section{Thermodynamics and Statistical Mechanics}

In my critique of Grimm's and de Regt's theories of understanding, I pointed out that physicists have disagreed on what counts as a satisfactory level of understanding. Some, like Newton or Einstein, were critical of certain received views and pushed for a deeper kind of understanding. In doing so, both noticed that the physics of their time did provide understanding-why but not a sufficient level of understanding-what or understanding-how. Newton's own theory could explain the falling apples and the motion of planets, but it was incapable of answering \emph{what} gravity is. Although in the \emph{Principia} Newton doesn't feign any hypotheses to a \emph{scientific description} of gravity, he seemed to be dissatisfied with this state of physics as he wrote in his famous letter to Bentley in 1693:
\begin{quote}
That gravity should be innate, inherent, and essential to matter, so that one body may act upon another at a distance through a vacuum, without the mediation of anything else, by and through which their action and force may be conveyed from one to another, is to me so great an absurdity that I believe no man who has in philosophical matters a competent faculty of thinking can ever fall into it. \citep[quoted in][p.\ 54]{Janiak:2008aa} 
\end{quote}
Newton himself was not able to deliver the ontological underpinning for gravity. Nevertheless, he agreed (with many of his Cartesian critics) that his theory of gravity was incomplete because it did not provide understanding-what nor understanding-how of gravitational phenomena.

Similarly, Einstein noticed shortcomings in understanding electromagnetic and quantum mechanical phenomena. \emph{What} is the nature of the electromagnetic field in giving rise to this asymmetry in explaining an electric current? This question lead Einstein to build special relativity, which delivered new scientific descriptions for space, time, light, and the electromagnetic field in general. In this way, special relativity gave us new ways of understanding \emph{how} nature works, and Einstein later did the same for gravity in his general theory of relativity. 

Taking what-questions seriously and demanding understanding-how, instead of mere understanding-why, Einstein also opened the door for revolutionary developments in quantum mechanics. The title of the EPR-paper shows this concern \citep*{Einstein:1935gf}: \emph{Can Quantum-Mechanical Description of Physical Reality Be Considered Complete?} They concluded that, given locality, quantum mechanics does not provide a complete description of reality. Even if quantum mechanics were complete in the sense of EPR, the wave-function  itself remains a mysterious object that cannot directly describe objects in three-dimensional space, since it is mathematically defined in configurations space. This prompted John \citet{Bell:2004ag} to introduce \emph{local beables} in order to mechanistically explain quantum phenomena that provide genuine understanding-how. I don't want to lose myself in the depths of quantum physics, as the significance of mechanisms and understanding-how for this theory is still heavily disputed (although I think the strongest arguments are indeed in favor of them). Instead, I want to rewind time a hundred years to discuss in detail my tri-partite distinction of understanding in thermodynamics and statistical mechanics. A lot can be learned and applied from this period to quantum physics and other (future) physical theories.

\subsection{Understanding in the Historical Development}
\label{subsec:thermo-history}
The modern form of thermodynamics was developed in the early and mid-1800s, although significant work has been done before already from the beginning of the 17\textsuperscript{th} century, when physicists and engineers began to build steam engines. The major goal was to make these engines efficient, and studies in thermodynamics helped in extracting more work from heat. The basic properties of a thermodynamic system are quantities, like volume, pressure, temperature, and entropy.\footnote{The concept of entropy found its way into thermodynamics only in 1854 in a paper by Rudolf Clausius (1822--1888), but it took another decade for Clausius to come up with the name ``entropy'' \citep[][p.\ 95]{Purrington:1997aa}.} These quantities are purely phenomenological, that is, they are measurable properties of macroscopic systems. The theory then gives three laws how thermodynamic systems behave. For all practical purposes, thermodynamics is all that is needed in order to build, use, and manipulate steam engines and explain thermodynamic behavior in general. 

The standard history of thermodynamics and statistical mechanics goes something like this. Physicists of the 19\textsuperscript{th} century didn't believe in atoms; thermodynamics was developed without reference to an underlying ontology; and Boltzmann was the lonely knight to defend the atomic theory against an army of positivists lead by Mach. The first two parts  are incorrect, and Boltzmann's fight against the rest mostly took place when Mach returned from Prague to Vienna in 1895---as we will see, another controversial part of Boltzmann's theory, besides atoms, was his use of probabilistic reasoning in physics.

Before the advent of modern thermodynamics, the caloric theory was the standard theory to account for the behavior of heat \citep{Fox:1971aa}. The most important justification for the caloric theory was provided by Joseph Black (1728--1799)  in 1757, when he discovered latent heat, that is, when adding heat to a system won't increase temperature, which occurs in phase transitions \citep[][p.\ 75]{Purrington:1997aa}. Black wasn't particularly interested in the scientific description of heat ('\emph{What} is heat?'), so it was Laplace and Lavoisier who advocated and developed the caloric theory and who explained latent heat by the behavior of this novel caloric fluid \citep[][p.\ 478]{Chang:2013aa}. The basic idea is that the caloric fluid binds with the particles. In a phase transition, the changes in this binding changes the phase without changing the temperature. Although Benjamin Thompson (later count Rumford) experimentally showed in 1798 that the caloric theory was untenable, by extracting a limitless amount of heat when boring cannons for the Bavarian monarchy, the caloric theory was still the predominant theory of heat until the 1820s. 

One important step toward the modern formulation of thermodynamics was taken in Sadi Car\-not's  little book \emph{Reflections on the Motive Power of Heat} (1824), in which Carnot (1796--1832) gave the first formulation of the second law of thermodynamics \citep[][p.\ 86]{Purrington:1997aa}: ``Wherever there exists a difference of temperature, wherever it has been possible for the equilibrium of the caloric to be re-established, it is possible to have also the production of impelling power'' \citep[][p.\ 48]{Carnot:1897aa}. Carnot was guided throughout his book by the caloric theory, which made it possible for him to imagine the flow of the caloric fluid in different physical set-ups. For example, in, what was later called, the Carnot cycle, Carnot ``made an explicit analogy to water engines, which went along nicely with his notion that caloric produced mechanical work in the course of ‘falling’ from a place of higher temperature to a place of lower temperature'' \citep[][pp.\ 497--8]{Chang:2013aa}. Even when it turned out that heat is not a special kind of fluid, this scientific description of heat made it possible for Carnot to make progress in thermodynamics and to find a new law. So apart from providing understanding-how, a scientific description can be used, even if it is wrong, to guide the intuition of a scientist to develop new physics.\footnote{This was also one of the main motivations for David Bohm to develop his own quantum theory, which he himself preferred to call the \emph{ontological interpretation}, although it was empirically equivalent to standard quantum mechanics \citep[see][for a detailed discussion]{Strien:2020aa}.}

After the death of Laplace in 1827, the caloric theory lost one of its most famous supporters; since then the atomic theory of matter gained more and more followers. Building on the results of Carnot and Joule, Clausius in 1850 gave a mathematical formulation of the first and second law of thermodynamics that is familiar to us today. This formulation of thermodynamics is only in terms of macroscopic phenomenological quantities, like temperature, volume, heat, and entropy. Although atoms are not part of Clausius' formulation of the laws of thermodynamics, Purrington remarks that a scientific description of matter in terms of atoms had been crucial to this development: 
 
 \begin{quote}
Neither Joule, Clausius, and Maxwell nor those who preceded them (namely, Waterston, Herapath, and König) could have developed their ideas without a thorough commitment to atoms or to molecules as collections of atoms. \citep[][p.\ 129]{Purrington:1997aa}
\end{quote}

In his other works, Clausius set out to explain properties of the specific heat of gases in terms of the motion of atoms, in particular, it was of great interest to explain the value of the quotient of the specific heat at constant pressure $c_p$ and at constant volume $c_v$ for air (and for other simple gases). The number $\frac{c_p}{c_v}=1.421$ for air couldn't be explained by macroscopic properties. The first to explain this value with the atomic theory was August Krönig (1822--1879) in 1856 aiming at understanding-how of this quotient. He  postulated only translational degrees of freedom for the motion of atoms, and calculated the theoretical value to be 1.67. This was too much off the mark. So Clausius introduced further internal degrees of freedom of atoms, and  these additional postulates about the mechanism behind specific heat cleared the way for Maxwell and Boltzmann to calculate the quotient of the specific heats, which yields 1.40 \citep[see][section 6.2, for further historic and mathematical details]{Regt:2017aa}.

As early as 1738, Daniel Bernoulli (1700--1782) used the atomic theory of matter in order to explain the pressure of a gas. In a similar spirit John Herapath (1790--1868) did so in 1820 and  John James Waterston (1811--1883) in 1845 \citep[][p.\ 767]{Darrigol:2013aa}. Their work, however, was not widely recognized, and only with the publications of Rudolf Clausius, probably because of his reputation as professor at the ETH Zurich, the atomic theory gained more traction in the physics community. In 1857, Clausius made the first step toward a mathematically precise atomic theory of gases by introducing the statistical concept of the \emph{mean free path}, which describes the average distance that atoms move before colliding. Building on Clausius work, James Clark Maxwell (1831--1879) used statistical arguments to derive for the first time in 1860 the velocity distribution of gas molecules in equilibrium, which now bears his name \citep[][pp.\ 769--70]{Darrigol:2013aa}: $v^2\exp{(-\frac{v^2}{T^2})}$, with $T$ the temperature of the gas. This was the time when the atomic theory of matter started to show its power to statistically explain the behavior of gases and to go beyond the laws of thermodynamics.

Ludwig Boltzmann, the most famous advocate of the kinetic theory of gas, built on Maxwell's work. In his papers of 1868 and 1871, he generalized the Maxwell distribution to atoms that have internal degrees of freedom and that could be affected by external forces, such that the distribution became $\exp{(-\frac{E}{kT})}$, with $E$ the sum of the kinetic and potential energy and $k$ the Boltzmann constant. He also proved that this generalized equilibrium distribution , which has been named the Maxwell--Boltzmann distribution, is stationary. In his landmark paper of 1872, Boltzmann accomplished several tasks at once. First, he derived (or rather conceptually justified) the Boltzmann equation, which describes how an arbitrary velocity distribution of atoms changes in time and ultimately approaches the Maxwell--Boltzmann distribution. Thus, the Maxwell--Boltzmann distribution is the \emph{unique} distribution for a gas in a box. In a next step, Boltzmann derived and interpreted the $H$-theorem (Boltzmann himself named it more appropriately the \emph{minimum theorem}). The $H$-theorem is the molecular analogue of Clausius' second law of thermodynamics, and Boltzmann thereby completed the reduction of thermodynamics to the atomic theory of matter. 

After criticism from Maxwell (in guise of his demon) and Loschmidt's reversibility objection, Boltzmann agreed that the $H$-theorem and therefore the second law of thermodynamics allows for exceptions, that is, not all systems in non-equilibrium will reach equilibrium, only most will do, as he noted in 1877:

\begin{quote}
One cannot prove that for every possible initial positions and velocities of the spheres, their distribution must become more uniform after a very long time; one can only prove that the number of initial states leading to a uniform state is infinitely larger than that of initial states leading to a non-uniform state after a given long time; in the latter case the distribution would again become uniform after an even longer time. \citep[quoted in][p.\ 775]{Darrigol:2013aa}
\end{quote}

In this period, although the kinetic theory could explain a lot, the adversaries against it began to rise due to the increasing popularity of positivism, especially in France, Germany, and Austria.  By the end of the 19\textsuperscript{th} century, Boltzmann was indeed basically alone defending the reality of atoms and the kinetic theory of gases. For many of his allies and sympathizers had died: James Clerk Maxwell ($\dagger$ 1879), Rudolf Clausius ($\dagger$ 1888), Josef Stefan ($\dagger$ 1893), and Johann Josef Loschmidt ($\dagger$ 1895). In 1895 after 28 years in Prague, Boltzmann's greatest rival and the main developer of positivism, Ernst Mach, returned to the University of Vienna, where Boltzmann started to teach in 1894. In contrast to Mach, who had young and accomplished followers, like  Friedrich Wilhelm Ostwald (1853--1932) and Max Planck (1858--1947), Boltzmann was not able to convince any rising star in the new generation.\footnote{Max Planck later changed his mind when he used Boltzmann's idea of quantizing energy for his work on blackbody radiation. Albert Einstein, although studying Boltzmann's work, was still too young a student at the ETH Zürich at this time.}


To my mind, a major reason why positivism in physics had gained popularity at the end of the 19\textsuperscript{th} century was that the macroscopic formulation of thermodynamics had been completed before the kinetic theory could deliver understanding-what and understanding-how of thermodynamical phenomena---this very situation plays a crucial role in the reception of quantum mechanics, too \citep{Cushing:1994aa}. There was at the beginning no agreement between the developers of the kinetic theory on what the precise scope of the theory was. Maxwell's demon and Loschmidt's reversibility objection showed that Boltzmann's initial understanding of the reduction of the second law of thermodynamics needed some polishing. 

This was the environment in which Mach could attract the young generation and some of the experienced physicists, like Gustav Kirchhoff (1824--1887) and Hermann von Helmholtz (1821--1894) \citep[see also][]{Holton:1992aa}, with phrases to be found at the end of his \emph{Conservation of Energy} (1872) and elsewhere:\footnote{Although often the same people worked also in electromagnetism, positivism didn't seem to be so prevalent in this theory. A rival theory to the field theory of electromagnetism was Wilhelm Weber's action-at-a-distance theory. It would be worthwhile to find out how this theory was influenced by positivism \citep[see, for instance,][]{Hon:2016aa}.}

\begin{quote}
The object of natural science is the connexion of phenomena, but theories are like dry leaves which fall away when they have ceased to be the lungs of the tree of science. \citep[quoted in][p.\ 85]{Blackmore:1972aa}\\

If we are astonished at the discovery that heat is motion, we are astonished at something which has never been discovered. \citep[quoted in][p.\ 86]{Blackmore:1972aa}
\end{quote}
Mach's doctrine was to abolish all non-observable entities from physical theories. Everything that cannot be directly confirmed by experiments was discredited as a mere form of fiction. Physics can and should only provide understanding-why; understanding-what and understanding-how are wishful thinking as they are based on unverifiable metaphysical claims. He saw in thermodynamics a theory that had fulfilled his ideas, and he ruthlessly dismissed anyone who tried to complete a macroscopic description by microscopic mechanisms. 

Ironically, because of Boltzmann's success in explaining the laws of thermodynamics the followers of Mach, especially Ostwald,  tried to develop a scientific description for thermodynamic systems. Instead of particles, for they cannot be directly observed, Ostwald chose energy as a kind of fluid as the underlying substance. Energy was said to be measurable and therefore preferable to particles \citep[][Sec.\ 11.4]{Cercignani:1998aa}. History showed where the research project of energetics has lead, but they did finally concede due to the success of Maxwell's and Boltzmann's work that it was worthwhile to investigate what matter is made of. The energetists in fact disagreed on what energy \emph{is} \citep{Deltete:2012aa}. For Ostwald, energy was indeed a substance of which matter was built, and he thereby conceded that some form of understanding-how is possible for some form of positivism. In contrast, the mathematical physicists Georg Helm (1851--1923) developed a version of energetics, in which energy is rather an abstract principle that unifies all physical phenomena. This doctrine was in this respect  closer to Mach's philosophy than Ostwald's. 

Finally, Ostwald admitted in the 1909 preface of his textbook \emph{Grundriss der allgemeinen Chemie} to had been on the wrong track:
\begin{quote}
I am convinced that we have recently come into possession of experimental proof of the discrete or grainy nature of matter, for which the atomic hypothesis had vainly sought for centuries, even millennia. \citep[quoted in][p.\ 209]{Cercignani:1998aa}
\end{quote}

Max Planck at first opposed the kinetic theory, too, because it made the ``universal'' laws of thermodynamics only valid in most cases; Planck couldn't accept that physical laws may allow for exceptions. In 1882, he wrote that ``the second law of the mechanical theory of heat is incompatible with the assumption of finite atoms [\dots] a variety of present signs seems to me to indicate that atomic theory, despite its great success, will ultimately have to be abandoned'' \citep[quoted in][p.\ 115]{Lindley:2001aa}. Having seen first-hand the great success of the atomic theory in his own work, Planck became one of the harshest critics of positivism. In 1904, he wrote 

\begin{quotation}
Though I am persuaded of the fact that Mach's system, pushed to all its logical consequences, does not contain any internal contradiction, I am no less convinced of the fact that this system has, after all, but a purely formal meaning: it is incapable of penetrating the very essence of science, and this because it is extraneous to the essential feature of any scientific research, i.e.\ the construction of a picture of the world which is rigorously stable, independent of the differences which mark the
generations and peoples. \citep[quoted in][p.\ 210]{Cercignani:1998aa}
\end{quotation}
So at the beginning of the 20\textsuperscript{th}, two of the harshest and most accomplished followers of Mach's positivism---both Planck and Ostwald received the Nobel Prize---reversed their earlier convictions and agreed to the reality of atoms, the merits of statistical mechanics, and that physics can indeed lead to a deeper form of understanding than previously believed by positivists. 

This historic excursion shows that physicists pursued different kinds of understanding of thermodynamics. Until the mid 19\textsuperscript{th} century, thermodynamics was always associated with a particular scientific description of matter. In the earliest beginnings in the 17\textsuperscript{th} century, physicists were searching for an atomic theory of gases due to Newton's influence. For a short period at the end of the 18\textsuperscript{th} century the phlogiston theory became popular until it was superseded by the caloric theory of Laplace and Lavoisier. After the fall of the caloric theory in the 1820s, the atomic theory and positivism filled the vacuum and battled for victory. Thanks to the ingenuity and perseverance of Ludwig Boltzmann and James Clerk Maxwell and others, thermodynamics did not remain in this positivistic state. They postulated that thermodynamic systems were composed of particles (understanding-what), whose motion guided by Newton's laws of motion determines the behavior of steam engines and gases (understanding-how). 

Mach, on the other hand, was not interested in scientific descriptions and hence in understanding-what. It was not even possible or scientific for him to ask what-questions in the first place. He aimed at understanding-why that only refers to the phenomenological level. The energetists, although Machian in spirit, were divided. Ostwald postulated energy as a substance (understanding-what), and one can argue that in doing so he indeed aimed at a mechanical explanation in Glennan's sense (and therefore at understanding-how).\footnote{I thank an anonymous reviewer for raising this idea.} Helm rather interpreted energy as a fundamental relational principle of nature and as a unifying principle for the laws of nature. Energy should explain the relations and changes between thermodynamic variables and ultimately between all physical variables. So it seems that energy played more of a nomological or dynamical role in Helm's philosophy as the cornerstone of understanding-why of physical phenomena. Like Mach, Helm refrained from providing further scientific descriptions of matter beyond what can be directly observed.

Both the energetists and the positivists had to concede the advantages of applying statistical mechanics to thermodynamics (and black-body radiation). They were ultimately swayed by the empirical success of statistical mechanics. But which kind of empirical success? I think they would not have supported statistical mechanics and the atomic theory if these were only able to recover the exact predictions of thermodynamics---something that has been explicitly made against the de Broglie--Bohm theory, too \citep{Becker:2018aa}---, and then the discussion would have focused rather on theoretical and philosophical differences between the two approaches (as it indeed happened in the early days of Maxwell's and Boltzmann's works on thermodynamics). As Planck did, they even criticized certain \emph{new} but unobservable predictions of statistical mechanics. It was only after a series of new predictions and applications of statistical mechanics (including Planck's works on black-body radiation and Einstein's works on Brownian motion) that statistical mechanics became widely accepted. I think that Maxwell's and Boltzmann's motivation for developing statistical mechanics did not only lie in making new predictions, but also in their metaphysical or epistemological requirement to discover what matter is ultimately made of. 

\subsection{Entropy: Clausius and Boltzmann}

After this historical case study, I want to illustrate my tri-partite distinction of understanding with a conceptual example, namely, two ways of defining entropy in thermodynamics and statistical mechanics according to Clausius and Boltzmann.\footnote{There are other definitions of entropy: the Gibbs entropy in statistical mechanics, the von Neumann entropy for quantum systems, and the Shannon entropy for information. \citet{Robertson:2021wa} shows how the Gibbs entropy relates to thermodynamics; \cite{Myrvold:2020ac} discusses how the von Neumann entropy relates to thermodynamics; and \citet{Carcassi:2021wc} do so for the Shannon entropy.} Changes in entropy are crucial in many thermodynamic processes and the different ways of defining entropy provide a different kind of understanding these very processes. One may argue that both the Clausius and the Boltzmann entropy provide us with understanding-how of thermodynamical processes, but I will show that the Clausius entropy does this to a much less satisfying degree and leaves important what-questions open, whose answers would give us a more fine-grained understanding-how by specifying a detailed mechanism.

\subsubsection{Clausius Entropy}
From the very beginning, thermodynamics has been  a phenomenological theory, whose primary domain was to elucidate the relationship between heat and work, more precisely, under which conditions heat can be transformed into work \citep{Lemons:2009aa,Lemons:2019aa,Myrvold:2020ac}. In order to quantify this conversion, Rudolf Clausius defined entropy \citep[][pp.\ 260--276]{Cardwell:1971aa}:

\begin{quote}
But as I hold it to be better to borrow terms for important magnitudes from the ancient languages, so that they may be adopted unchanged in all modern languages, I propose to call the magnitude $S$ the \emph{entropy} of the body, from the Greek word $\tau\rho o\pi\eta$, \emph{transformation}. I have intentionally formed the word \emph{entropy} so as to be as similar as possible to the word \emph{energy}; for the two magnitudes to be denoted by these words are so nearly allied in their physical meanings, that a certain similarity in designation appears to be desirable. \citep[][p.\ 357]{Clausius:1867aa}
\end{quote}
Although Clausius's mathematical definition of entropy hinges on details of the Carnot cycle and Thomson's definition of absolute temperature \citep{Thomson:1853tc,Smith:1976vh}, the general idea is that entropy is a conserved quantity in a reversible\footnote{It suffices for our purposes to characterize  reversible processes as minimally (that is, quasi-statically) changing the physical system so that they can be conducted in the reverse order \citep[for a detailed discussion, see][]{Norton:2016ab}.} cyclic\footnote{A thermodynamic process is called cyclic, if it returns to its initial state \citep[][p.\ 48]{Norton:2016ab}.} thermodynamic process. Consider, for example, the flow of heat $Q$ between two heat reservoirs at temperature $T_C$ and $T_H$, respectively, with $T_C<T_H$ \citep[][Ch.\ 10]{Lemons:2019aa}. What is the entropy change of the entire system, after the heat $Q$ flowed from the hotter to the colder reservoir? This is, indeed, an irreversible process because there is no way to directly reverse the flow of heat in the other direction. Nevertheless, we can replace this irreversible process by a reversible one:
\begin{enumerate}
\item[i)]
We separate the two reservoirs and use a box of gas with a moving piston to extract the heat $Q$ from the hot body at $T_H$. The gas will thereby expand and push the piston until it reaches the temperature $T_H$.
\item[ii)]
We then separate the box of gas and let the gas adiabatically\footnote{An adiabatic process is a process in which the physical system is closed, that is, it does not exchange heat or matter with an outside system. The word \emph{adiabatic} comes from the Greek word \emph{adiábatos}, which literally means impassable (from a ``not'' + diabatós ``to be crossed'').} expand until its temperature drops to $T_C$.
\item[iii)]
We bring the box of gas in thermal contact with the colder reservoir at $T_C$ and push the piston until the heat $Q$ gets transferred into the reservoir. 
\end{enumerate} 

In this reversible process, the heat $Q$ got first transferred \emph{out} of the reservoir at temperature $T_H$ and then transferred \emph{into} the other reservoir at temperature $T_C$. According to Clausius's definition of entropy, the hotter reservoir lost $\frac{Q}{T_H}$ of entropy and the colder one gained $\frac{Q}{T_C}$, such that the net change in entropy is 
$$
\Delta S=Q\left( -\frac{1}{T_H}+\frac{1}{T_C} \right)>0.
$$
Although reversible, the entropy does not remain constant in this process, because the gas in the box does not return to its initial state---the piston will not reach the same level as at the beginning. In general, Clausius showed that the net entropy for any cyclical process is
$$
\oint \frac{\delta Q}{T}\leq 0.
$$
The integral sums up all the incremental changes in entropy over any cyclical process. If the process is reversible, equality holds; if the process is irreversible, the net result is negative, which means that the final state has higher entropy than the initial state---given the proper sign convention as we can see in the above example of a reversible (non-cyclic) process. This inequality is Clausius's formulation of the \emph{2\textsuperscript{nd} Law of Thermodynamics} \citep[][section 3 and 4]{Robertson:2021wa}. 

This law describes how entropy changes in a thermodynamics process. It does not explain, however, \emph{why} a thermodynamic system that is forced out of equilibrium will seek spontaneously by itself a new state of equilibrium.\footnote{Thermodynamic equilibrium is defined to be a state in which the macroscopic variables of a system, like pressure, volume, temperature, and entropy, do not change over (sufficiently long) time \citep[][section 3.1]{Robertson:2021wa}.} For example, the 2\textsuperscript{nd} Law of Thermodynamics does not say why heat spontaneously flows from a hot to a cold body or why the gas in a box expands once in thermal contact with a heat reservoir. This spontaneous behavior is presupposed in \emph{all} thermodynamic processes. Since it is so fundamental, \citet[][p.\ 528]{Brown:2001aa} called it the \emph{Minus-First Law of Thermodynamics} \citep[see also,][]{Myrvold:2019ab}: \emph{An isolated system in an arbitrary initial state within a finite fixed volume will spontaneously attain a unique state of equilibrium}.

We can ask two questions about the heat flow from a hotter to a colder body:
\begin{enumerate}
\item
Why does a system in non-equilibrium finally attain a state of equilibrium?
\item
How does this system change its entropy between two equilibrium states?
\end{enumerate}
The first question is answered by the minus-first law: the system just does it by itself. No further explanation is given for \emph{how} the system does it. This is a pure case of understanding-why without any further relation to the micro-constituents of the system. 

The second question is answered by the 2\textsuperscript{nd} Law (in conjunction with the -1\textsuperscript{st}). Thermodynamics postulates that the system has a certain temperature and that heat flows either directly between the hotter and the colder body or (factually or hypothetically) between the heat reservoirs and the box of gas. Hence, thermodynamics arguably provides us with a certain degree of understanding-how, as it postulates temperature and the flow of heat. Especially, the flow of heat may remind us of some mechanism that does explain how temperature or pressure changes. Heat as described in thermodynamics, however, is not a mechanism in Glennan's sense because thermodynamics is silent about what composes heat. Heat exists as something that is not directly observable but indirectly measureable and is exchanged in certain processes, but there is no scientific description of heat. In contrast to the first question, the second question does give us some form of understanding-how---some ``proto-understanding-how''---, and yet it is obvious that more can and should be said more about the nature of heat---which is in fact accomplished by statistical mechanics. So with regard to understanding, we learn two things from thermodynamics:
\begin{enumerate}
\item[A.]
Thermodynamic quantities, like heat, temperature, and entropy, are macroscopic phenomenological quantities. They are defined to be measurable quantities without a scientific description of what they actually are in terms of the micro-constituents of a physical system. Therefore, thermodynamic quantities can be construed as functional because they are defined by what they do rather than by what they are \citep[][section 2]{Robertson:2021wa}.
\item[B.]
Thermodynamic behavior is explained by the change of these macroscopic variables. In particular, the change from non-equilibrium to equilibrium is explained by a fundamental law, the Minus-First Law, that postulates this very phenomenon. Therefore, this law only provides understanding-why for thermodynamic processes (without understanding-what or understanding-how), whereas the 2\textsuperscript{nd} Law provides some incomplete understanding-how, as heat, if interpreted as some form of stuff, is postulated to exist and exchanged in certain processes.
\end{enumerate}
Statistical mechanics will complete thermodynamics and render genuine understanding-how for why and how thermodynamic systems behave. A crucial ingredient is Boltzmann's definition of entropy.

\subsubsection{Boltzmann Entropy}

\begin{figure}[ht]
\centering
\includegraphics[width=8cm]{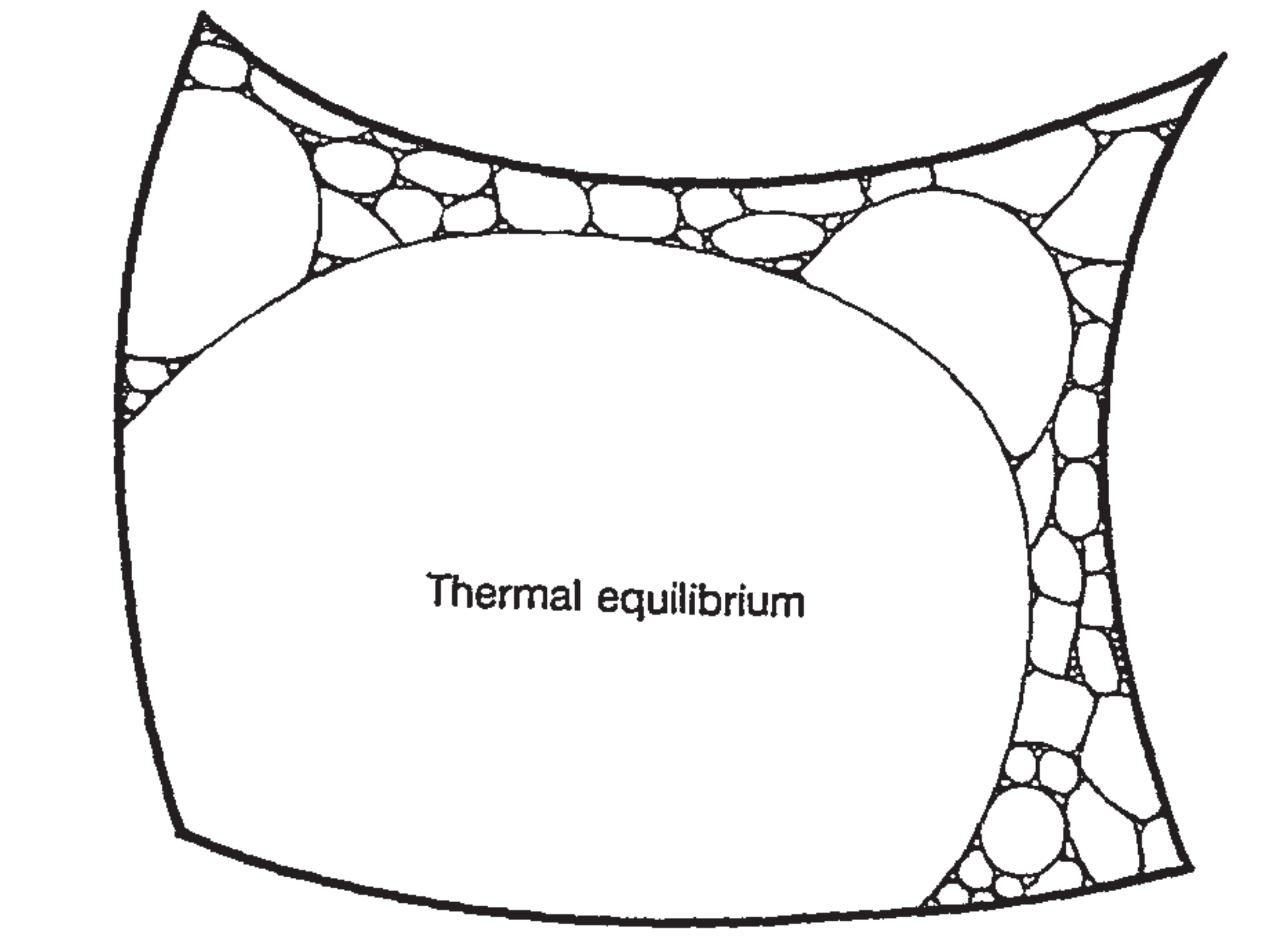}
\caption{Clusters in phase space according to thermodynamic macrostates and the appropriate measure, as depicted by Roger \citet[][p.\ 402]{Penrose:1989aa}. Thermal equilibrium is by far the largest macrostate.}
\label{fig:phase-space-cluster}
\end{figure}

In his reduction of thermodynamics to statistical mechanics, Boltzmann answered the two questions that have been left open in thermodynamics \citep[][for a comprehensive analysis of Boltzmann's papers]{Darrigol:2018aa}:
\begin{enumerate}
\item
\emph{What} is entropy (and all the other thermodynamic quantities)?
\item
\emph{How} do thermodynamic systems approach a (unique) state of equilibrium?
\end{enumerate}
By answering the first question, Boltzmann had the tools to answer the second. As we said in section \ref{subsec:thermo-history}, Boltzmann postulated and justified the existence of atoms such that every thermodynamic system consists of $N$ particles, where $N$ is of the order of Avogadro's constant, that is, approximately $10^{23}$. Since a particle is completely described by its position $\vec{x}$ and momentum $\vec{p}$, we can summarize the complete physical state of $N$ particles as $\left( \vec{x}_1,\vec{p}_1,\vec{x}_2,\vec{p}_2,\dots,\vec{x}_N,\vec{p}_N\right)$, which is called a microstate. All these points taken together comprise phase space, which has roughly $6\times10^{23}$ dimensions. In order to get macrostates, one needs to partition phase space $\mathcal{P}$ (see Figure \ref{fig:phase-space-cluster}). A macrostate of a microstate arises from a map $M$ that assigns to every microstate $X$ a macrostate $M(X)$ corresponding to one of the subsets $\mathcal{P}_M\subseteq \mathcal{P}$ according to the partition: $M(X)$ is the macrostate of $X$ if $X\in\mathcal{P}_M$.

The tool that ultimately leads to explain the approach to equilibrium is Boltzmann's definition of entropy assigned to every point in phase space:

\begin{equation}
\label{eq:entropy}
S_B(X):=k_B\ln \lvert \mathcal{P}_{M(X)}\rvert, 
\end{equation}
where $k_B$ is Boltzmann's constant and $\ln$ is the natural logarithm.\footnote{This definition of entropy is written on Boltzmann's tomb, although Boltzmann actually defined entropy in a different mathematical way, even if similar in spirit. This formula for entropy was due to Max Planck \citep[][p.\ 783]{Darrigol:2013aa} and taken up as the foundation for the neo-Boltzmannian project of statistical mechanics \citep{Callender:1999uf,Goldstein:2001aa,Lebowitz:1993aa,Lebowitz:1993ac,Lebowitz:1994aa,Lebowitz:2008aa,Penrose:1989aa}.} The main part of Boltzmann's entropy is $\lvert \mathcal{P}_{M(X)}\rvert$, which deserves some elaboration. In order to measure the sizes of the subsets $\mathcal{P}_{M(X)}$, one needs to introduce a measure $\lambda$, which assigns a number to every such subset. Conventionally, if the system is finite, one normalizes the measure to $1$ such that the size of the entire phase space would be $1$. In the entropy formula,  $\lvert \mathcal{P}_{M(X)}\rvert$ denotes the size of $\mathcal{P}_{M(X)}$ according to the appropriate measure $\lambda$, that is, $\lvert \mathcal{P}_{M(X)}\rvert=\lambda \left( \mathcal{P}_{M(X)}\right)$. The measure $\lambda$ is conventionally interpreted as a probability measure.\footnote{There is a debate on whether one should rather interpret the measure as a \emph{typicality measure}, but this should not concern us here \citep[see, e.g.,][]{Volchan:2007aa,Frigg:2009aa,Werndl:2013aa,Lazarovici:2015aa,Myrvold:2019ab,Maudlin:2019ab,Hubert:2020aa}.} Since for real physical systems, like heat flowing from a hot to a cold body or gases in a box, the phase space volume of thermal equilibrium has by far the largest volume according to the probability measure, it has the highest probability for a system to be in or to approach it.  

What is entropy then, and how does Boltzmann's idea improve on Clausius's? Notice that, unlike thermodynamics, statistical mechanics starts with a scientific description of physical systems on the fundamental level: all thermodynamic systems are composed of particles and all thermodynamic quantities supervene on the configuration, the properties, and behavior of these particles. In this way, statistical mechanics gives us a complete understanding of \emph{what} thermodynamic systems are composed of (understanding-what). Entropy is a measure of the number of microstates (that is, of states of the underlying mechanism) that give rise to the same macrostate (that is, the same macroscopic quantities). In particular, the mysterious quantity heat and also temperature in the Clausius entropy are described in terms of particles: heat flows when particles are exchanged or interact, and temperature is a form of average kinetic and potential energies of particles.

How do thermodynamic systems approach a (unique) state of equilibrium according to statistical mechanics? If we put two heat reservoirs into thermal contact, the entire system is in non-equilibrium. Because the equilibrium state has such a large phase space volume, the probability that the system will reach this state (rather quickly) is close to 1. 
In other words, if we zoom into the phase space region of this low entropy macrostate, almost all microstates will move to a macrostate with higher entropy and ultimately to equilibrium. It is physically possible that a low entropy macrostate goes into another low entropy macrostate, but very few microstates within this macro-region do that. According to Boltzmann's reduction of thermodynamics to statistical mechanics, it is physically possible that heat spontaneously flows from a colder to a hotter body, but the probability is so low that we won't actually observe such behavior. In this way, Boltzmann provided a statistical-mechanical explanation, and therefore a statistical way of understanding-how, of thermodynamic processes and the laws of thermodynamics, in particular the  -1\textsuperscript{st} (minus-first) and the 2\textsuperscript{nd} Law of Thermodynamics.

Statistical mechanics specifies a concrete mechanism for thermodynamic systems, namely, particles in motion that interact through classical forces and follow Newton's laws---Boltzmann's statistical reasoning supervenes on this mechanism. Therefore, thermodynamic processes can also be causal-mechanically explained: given the microstate of a thermodynamic system, Newton's laws of motion determine the future behavior of the system. This is another way of  understanding-how of thermodynamic processes.\footnote{One may think that because of the imaginary ensembles Gibbsian statistical mechanics (which is the most popular formulation among physicists) does not specify a mechanism---I thank an anonymous reviewer for raising this concern. First, \citet{Robertson:2021wa} argues that ensembles are not essential for Gibbsian statistical mechanics; they are rather introduced to interpret the probabilities as hypothetical frequencies. Second, Gibbs introduces a probability measure over sets of microstates, which specify the probability in which particular microstate the system is. In this sense, Gibbs presupposes a mechanism or, at least, the existence of microstates. It seems, however, that a Newtonian mechanism does not play such a prominent role for Gibbs as it does for Boltzmann, whose entire project is to \emph{derive} statistical mechanics from Newtonian physics. Since Gibbs follows a more pragmatic approach than Boltzmann, \citet{Frigg:2019aa} call Gibbsian statistical mechanics an \emph{effective theory} and Boltzmannian statistical mechanics a \emph{fundamental theory}.} Statistical mechanics gives us a deeper understanding for the same processes that are explained by thermodynamics (apart from explaining new phenomena, like mixing). Statistical mechanics is, therefore, a good example of how to transform a theory that renders only understanding-why or an incomplete form of understanding-how into a theory providing us with genuine understanding-what and understanding-how.

\section{Conclusion}
I argued that scientific inquiry, and therefore our quest for understanding nature, does not only start by asking \emph{why} but also by asking \emph{what}. Answers to what-questions are scientific descriptions, which are valuable by themselves because they fulfill a basic epistemic need. With the help of scientific descriptions I introduced three kinds of understanding: understanding-what, understanding-why, and understanding-how. Current theories of understanding, especially the theories by Grimm and de Regt, have focused on understanding-why. I argued that understanding-how is a deeper kind of understanding than merely understanding-why, because it requires, in addition, to understand what something is made of and how these constituents generate or produce the phenomenon. Understanding-how is provided by mechanistic explanations. 

One may object that what-questions are primarily metaphysical questions and therefore not the domain of physicists. But what-questions are common place, especially in chemistry and biology, and also in physics \citep{Chakravartty:2017wb}. I side with \citet{Felline:2019aa} that what a physical system is made of is primarily a scientific question and not merely metaphysics (although metaphysics can fill in some important details and add to a more comprehensive picture of the world ); the overall postulation is a scientific endeavor, that ought to be provided with a physical theory. 

We also saw in the discussion of thermodynamics that a theory of understanding touches on the realism-antirealism debate.\footnote{I thank an anonymous reviewer for pushing me on this point.} I think that there is some overlap between these areas (which would be also interesting to explore further), but they start from two different questions. Whereas the realist or anti-realist asks,``What does a theory tell us about the world and what is the right attitude of scientists toward scientific theories?'' a theory of understanding asks, ``What are criteria to understand the world and what can we expect to understand in the first place?'' The realism-antirealism debate focuses more on the metaphysical consequences and the postulation of unobservable entities within scientific theories.  The understanding debate, on the other hand, focuses more on the epistemic and pragmatic sides of scientific inquiry and shows different ways scientists try to understand the world. 
I think that understanding starts from a more general point of view and focuses on epistemic criteria, wishes, and demands for what we can aim to know about the world in the first place. Grimm, for example, begins his investigation into the nature of understanding from this high vantage point, and, as we saw, de Regt considers also other representational devices for scientific inquiry apart from scientific theories.

Although both fields start with different questions, they do meet on certain issues and are connected in important respects.
 When we distinguish different kinds or levels of understanding we do quickly step into the waters of the realism-antirealism debate. Scientific descriptions and mechanisms deepen our understanding and fulfill an epistemic need to know what something is made of and how it behaves, but, the anti-realist may ask, ``Shall we \emph{believe} in these entities? How certain are we that the world matches these descriptions?'' Roughly, understanding emphasizes more how a scientist relates to a theory, whereas the realism-antirealism emphasizes more how a theory is related to the world, how the theory properly represents the world, and on which grounds one can and should believe in the theory. As \citet{Regt:2016aa} argue, and I agree with that, a certain degree of understanding can be even accomplished when the theory is wrong, that is, when the theory does incorrectly represent the world. This would be, of course, a no-go for a realist and rather fuel the anti-realist's skepticism that current or future theories do not match the unobservable ontology of the world. The significance of the pessimistic meta-induction seems to be, therefore, more important for the realism-antirealism debate than for theories of understanding. 
 
One can also see the trends within the realism-antirealism debate in the theories of understanding. Both Grimm's and de Regt's theories reflect the antirealist's tendencies in past and current physical theories, in particular the trends in quantum mechanics. Both agree that physics in the antirealist or positivistic tradition provides understanding enough as long as the scientific community regards these standards as sufficient. Both do not want to establish prescriptive norms that would push scientists to improve their degree of understanding. My account, on the other hand, builds on the realist's attitude for developing standards of scientific understanding. My historic case studies showed that even in times where antirealism was the received framework of the majority of physicists, some, like Einstein, Boltzmann, and   Bell, did dissent from this norm and demanded a deeper form of understanding. I, therefore, believe in an optimistic meta-induction: since these demands for scientific descriptions and understanding-how were so successful, one should raise them as general norms for understanding past, current, and future physics.

I hope that we can develop theories of scientific descriptions as has been done for scientific explanations and that a closer focus on what-questions influence philosophical analysis of physics (and science in general), as well as guide how physics is practiced and how physical theories can be constructed---given the historical success of statistical mechanics, for example. I hope philosophers and physicists come to better appreciate the importance of scientific descriptions. This may then lead to a more fine-grained and more comprehensive hierarchy of understanding that considers different kinds of descriptions, explanations, and theories, as well as pragmatic aspects of scientific practice.

\section*{Acknowledgements}
I wish to thank Frederick Eberhardt, Christopher Hitchcock, and Tim Maudlin for their support and many helpful discussions. I also thank the audience of the \emph{Caltech Philosophy of Physics Reading Group} for their comments. I also thank two anonymous reviewers for helpful comments. Special thanks go to Joshua Eisenthal and Maaneli Derakhshani for their help on the passages on Einstein and to Charles Sebens and Federica Malfatti for thoroughly reading and commenting on previous drafts of this paper. Funding for this research was  partially provided by the Swiss National Science Foundation as part of the Early Postdoc.Mobility Fellowship, grant no.\ 174745.

\bibliographystyle{abbrvnat}
\bibliography{references}
\end{document}